\documentclass{webofc}

\usepackage[varg]{txfonts}   
\usepackage{hyperref}
\usepackage{url}
\hypersetup{colorlinks=true,citecolor=blue,urlcolor=blue,linkcolor=blue}
\newcommand{\GeV}{\ensuremath{\mathrm{GeV}}}
\newcommand{\MeV}{\ensuremath{\mathrm{MeV}}}
\newcommand{\fm}{\ensuremath{\mathrm{fm}}}
\usepackage{xcolor}

\usepackage{comment}

\usepackage{ulem}
\newcommand{\Remove}[1]{{\color[rgb]{0.8,0.3,0.3}\ifmmode\text{\sout{$#1$}}\else\sout{#1}\fi}}

\begin{document}
\title{$\Lambda$ and $\Sigma$ potentials in dense matter based on chiral EFT:
Bridging heavy-ion collisions, hypernuclei, and neutron stars}
%
%

\author{\firstname{Asanosuke} \lastname{Jinno}\inst{1}\fnsep\thanks{\email{jinno@ruby.scphys.kyoto-u.ac.jp}} \and
        \firstname{Koichi} \lastname{Murase}\inst{2} \and
        \firstname{Yasushi} \lastname{Nara}\inst{3} \and
		\firstname{Johann} \lastname{Haidenbauer}\inst{4}
}

\institute{Department of Physics, Faculty of Science, Kyoto University,
Kyoto, 606-8502, Japan
\and
           Department of Physics, Tokyo Metropolitan University,
Hachioji 192-0397, Japan
\and
           Akita International University,
 Yuwa, Akita-city 010-1292, Japa
\and
		   Institute for Advanced Simulation, 
Forschungszentrum J\"ulich, D-52425 J\"ulich, Germany
          }

\abstract{
	The $\Lambda$ and $\Sigma$ directed flows at $\sqrt{s_{NN}}=4.5~\GeV$
	are investigated to examine their sensitivity to the hyperon single-particle potentials.
	The single-particle potentials are obtained from $G$-matrix calculations
    with two- and three-body forces based on SU(3) chiral effective field theory.
	The $\Lambda+\Sigma^0$ directed flow shows
	sensitivity to the variation in the $\Sigma$ single-particle potential.
	Its effect is more pronounced for the $\Sigma^0$ directed flow.
}
\maketitle
\section{Introduction}
\label{intro}

A unified approach relating heavy-ion collisions, low-energy
nuclear experiments, and neutron stars with the modern nuclear
force derived from chiral effective field theory (EFT) has
provided valuable insights into the properties of the equation
of state of dense matter (e.g. Ref.~\cite{Huth:2021bsp}).
For a microscopic description of dense matter,
such an approach must be extended to the hyperon sector.
A precise understanding of the hyperon
single-particle potentials in nuclear matter is essential
for addressing the long-standing hyperon puzzle in neutron stars.

In this contribution, we examine the $\Lambda$ and $\Sigma$
directed flows at $\sqrt{s_{NN}}=4.5~\GeV$ and
investigate their sensitivity to the hyperon single-particle potentials~\cite{Gerstung:2020ktv},
obtained from the $G$-matrix calculations with two- (2BF) and three-body forces (3BF)
based on chiral EFT\@.
These potentials are sufficiently repulsive
at high densities to solve the hyperon puzzle and have been shown
to be consistent with the hypernuclear spectroscopy~\cite{Jinno:2023xjr}.
We employ a relativistic quantum molecular dynamics (RQMD) model
implemented into the Monte-Carlo event generator
\texttt{JAM2}\footnote{\url{https://gitlab.com/transportmodel/jam2}}
with updates since our previous work~\cite{Nara:2022kbb}.

As novel feature, we implement the $\Sigma$ single-particle potential,
evaluated from the same hyperon-nucleon (YN) interaction as that of the $\Lambda$.
In our previous work~\cite{Nara:2022kbb} on the $\Lambda$ directed flow,
all hyperons and their resonances were assumed to feel the same single-particle
potential as the $\Lambda$. Meanwhile, the $\Sigma^0$ contribution should have an impact
because the $\Lambda$ directed flow is in essence the $\Lambda + \Sigma^0$
directed flow due to the process $\Sigma^0 \rightarrow \gamma + \Lambda$.
Also, the $\Sigma$ single-particle potential is expected to be
more repulsive compared to that of $\Lambda$ as reflected in
the non-existence of $\Sigma$ hypernuclei except for ${}^4_\Sigma \mathrm{He}$.
We discuss the influence of the $\Sigma$ single-particle potential
on the $\Lambda+\Sigma^0$ and the $\Sigma^0$ directed flows.

\section{$\Lambda$ and $\Sigma$ single-particle potentials from chiral EFT}
\label{sec:U_chi}

We use the Lorentz-vector version of the relativistic
quantum molecular dynamics (RQMDv) model~\cite{Nara:2021fuu}
implemented in \texttt{JAM2},
which gives the rapidity dependence
of the proton directed flow compatible with the data
for a wide range of beam energies $\sqrt{s_{NN}}=3$--$20~\GeV$~\cite{Nara:2021fuu}.

We employ the $\Lambda$ and $\Sigma$ potentials evaluated from
2BFs and 3BFs based on $\mathrm{SU(3)}$ chiral EFT\@.
The $G$-matrix calculation is employed to calculate
the single-particle potentials.
The 2BF is chosen to be one of the chiral potential from the J\"ulich-Bonn group,
NLO13~\cite{Haidenbauer:2013oca} with the cutoff of $500~\MeV$\@.
The 3BF is implemented as an effective density-dependent 2BF
in which the low-energy constants (LECs) are estimated
via the decuplet-saturation assumption~\cite{Petschauer:2016pbn,Gerstung:2020ktv}.
The LECs involved in the 3BF are determined
to fulfill two constraints:
$U_\Lambda(\rho_0,k=0) \approx -30~\MeV$~\cite{Gal:2016boi} from the $\Lambda$ hypernuclear spectroscopy and
$U_\Lambda(3\rho_0,k=0)>80~\MeV$ in pure neutron matter
so that the appearance of $\Lambda$'s in neutron stars is suppressed.
We choose the set of the 3BF LECs that
produces the most repulsive $\Sigma$ potential.

The $\Lambda$ and $\Sigma$ single-particle potentials are
implemented in RQMDv by fitting the following forms
to the results from chiral EFT:
\begin{gather}
	U(\rho,k) = U_\rho(\rho/\rho_0) + U^0_m (\rho,k), \\
	U_\rho(u) = au + bu^{4/3} + cu^{5/3},\quad
	U^0_m(\rho(x),k) = \dfrac{C}{\rho_0} \int d^3 k' \dfrac{f(x,k')}{1+\left[(\boldsymbol{k}-\boldsymbol{k}')/\mu\right]^2},
\end{gather}
where $a$, $b$, $c$, $C$ and $\mu$ are fitting parameters
and $f(x,k)$ is the baryon single-particle distribution function.
In the actual heavy-ion simulation, we implement the
momentum-dependent potential in the Lorentz-vector form
$U^\mu_m$~\cite{Nara:2022kbb}.

\begin{figure}[h]
\centering
\includegraphics[width=4.5cm,clip]{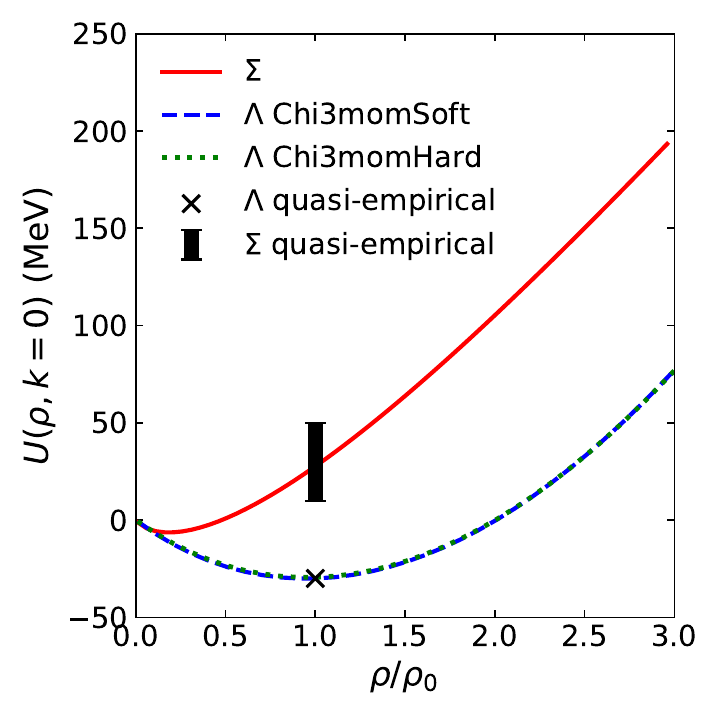}
\includegraphics[width=5cm,clip]{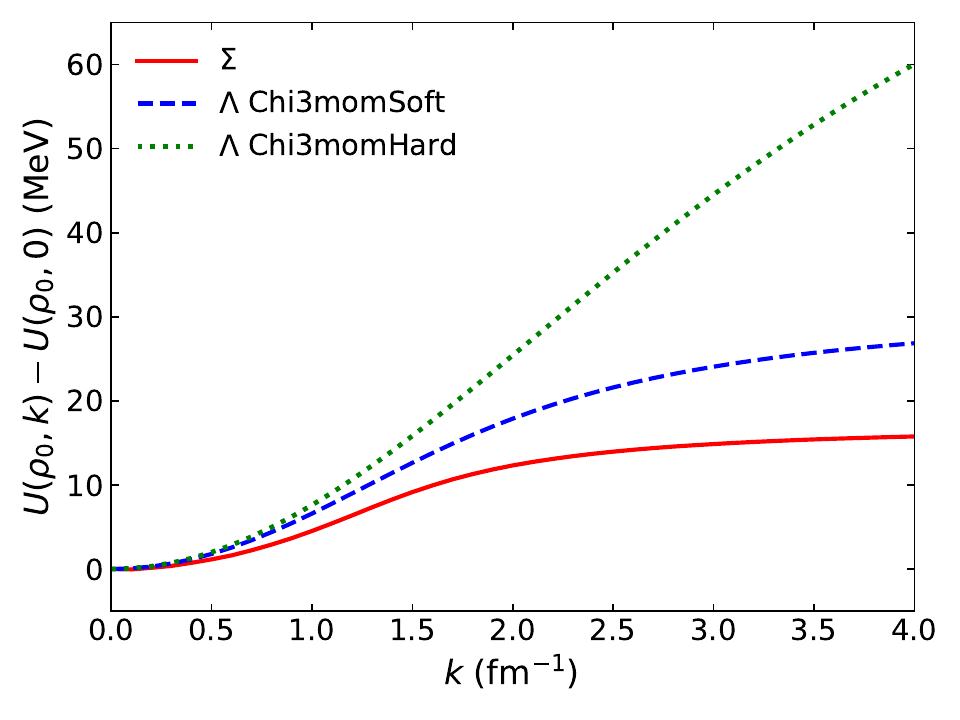}
\caption{Density dependence (left panel) and momentum dependence (right panel) of
the $\Lambda$ and $\Sigma$ single-particle potentials in symmetric nuclear matter.
The momentum dependence is subtracted by its value at $k=0$.
The single-particle potentials are fitted to the results obtained
for two- and three-body forces based on chiral EFT\@.
The solid lines correspond to the $\Sigma$ single-particle potential.
The dashed and dotted lines represent the $\Lambda$ single-particle potentials
with different momentum dependencies above $k>1~\fm^{-1}$.
The quasi-empirical values at $\rho_0$ are taken from Ref.~\cite{Gal:2016boi}.
}
\label{fig:pot}       
\end{figure}

The fitted results for the density dependence of
the hyperon single-particle potentials are shown on
the left panel of Fig.~\ref{fig:pot}. The $\Sigma$ single-particle potential
is much more repulsive compared to the $\Lambda$ single-particle potential.
This behavior is consistent with the quasi-empirical value inferred
from $\Sigma^-$ atoms and $(\pi^+,K^+)$ inclusive
spectra: $U_\Sigma(\rho_0,k=0)=30\pm 20~\MeV$~\cite{Gal:2016boi}.

For the momentum dependence, we prepared two scenarios to 
simulate the uncertainty in chiral EFT\@.
Chi3momHard and Chi3momSoft are
constructed to reproduce the chiral EFT
result~\cite{Kohno:2018gby} up to $k=2.5~\text{fm}^{-1}$ and up to
$1.0~\text{fm}^{-1}$, respectively, as shown on the right panel of Fig.~\ref{fig:pot}.
The momenta correspond roughly to the potential cutoff of $\approx 500~\MeV$
and to 40\% of its value.
The $\Sigma$ momentum dependence is constructed
by a similar procedure as Chi3momSoft.

\section{$\Lambda$ and $\Sigma$ directed flows}
\label{sec:v1}

We consider the rapidity dependence of the hyperon directed flows,
\begin{equation}
v_1=\langle \cos \phi\rangle
=\Biggl\langle \frac{p_x}{\sqrt{p_x^2+p_y^2}}\Biggr\rangle,
\end{equation}
where $\phi$ is the azimuthal
angle measured from the reaction plane, and $p_x$ and $p_y$
are the transverse momenta of a particle.
The brackets indicate averaging over all events and particles.

Let us mention three updates in \texttt{JAM2}
since our previous work~\cite{Nara:2022kbb}:
First, the collision term in the Boltzmann equation
is updated to the Poincar\'e covariant one~\cite{Nara:2023vrq}.
Second, the hyperon-nucleon cross section is updated
to the ones calculated from the up-to-date chiral force at
next-to-next-to-leading order~\cite{Haidenbauer:2023qhf}.
Third, a new RQMD equation of motion~\cite{Nara:2025pkg}
is employed, which provides an accurate simulation of the equation of state,
in contrast with the traditional QMD calculations~\cite{TMEP:2021ljz,TMEP:2023ifw}.
We have confirmed that these updates have only a subtle impact
on the hyperon directed flows.
We label the results from the version used in Ref.~\cite{Nara:2022kbb} as RQMDv1 and
the new one as RQMDv2.

\begin{figure}[h]
\centering
\includegraphics[width=5cm,clip]{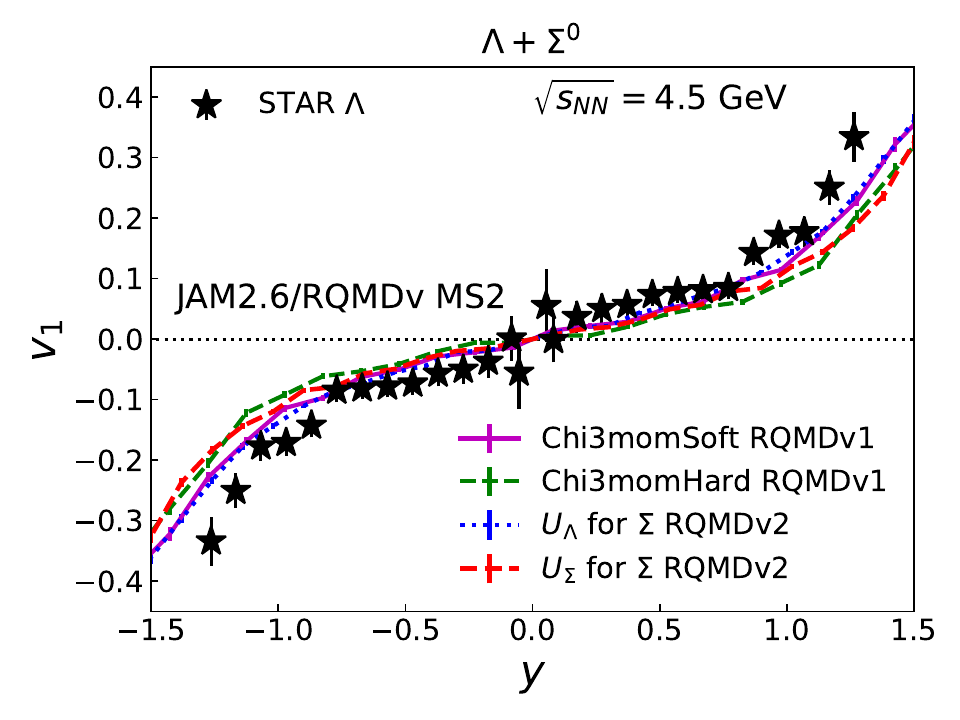}
\includegraphics[width=5cm,clip]{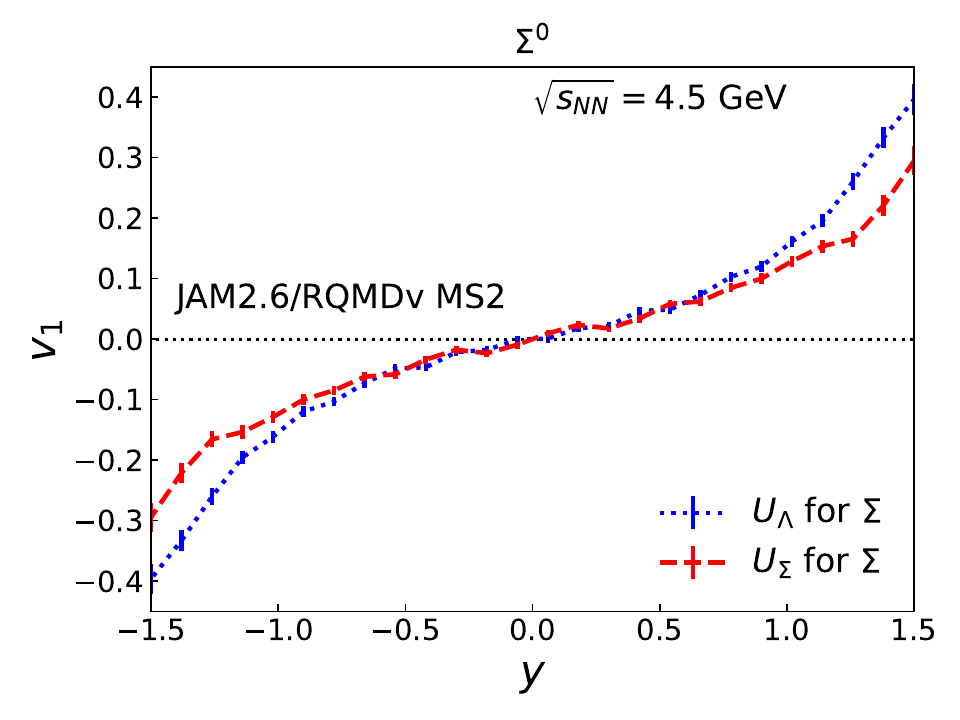}
\caption{Directed flows of $\Lambda+\Sigma^0$ (left panel) and
$\Sigma^0$ (right panel) as functions of the rapidity at
$\sqrt{s_{NN}}=4.5~\GeV$
in the mid-central $\mathrm{Au}+\mathrm{Au}$ collisions.
The STAR data is taken from Ref.~\cite{STAR:2020dav}.
}
\label{fig:v1}
\end{figure}

The results of the $\Lambda+\Sigma^0$ directed flow in
mid-central $\mathrm{Au} + \mathrm{Au}$ collisions at $\sqrt{s_{NN}}=4.5~\text{GeV}$
are shown on the left panel of Fig.~\ref{fig:v1} and
compared with the STAR data~\cite{STAR:2020dav}.
In the RQMDv1 calculations, a suppression is found in case of the
harder momentum dependence, as discussed in our previous
study~\cite{Nara:2022kbb}.
This illustrates that the harder momentum dependence leads
to more repulsive hyperon potentials.
Experimental information on the optical potential of $\Lambda$
would be useful for reducing the model uncertainty.
Substituting the $\Sigma$ single-particle potential for the $\Lambda$
one results in a similar suppression.

The effect of the $\Sigma$ potential is more significant for the $\Sigma^0$
directed flow than for the $\Lambda+\Sigma^0$ case, where the particle production
ratio of $\Lambda/\Sigma_0$ is $R = 3.05$ at $\sqrt{s_{NN}}=4.5~\GeV$,
as simulated in \texttt{JAM2}.
We note that the HADES experiment~\cite{Becker:2024nqu}
at $\sqrt{s_{NN}}=2.55~\GeV$ report a ratio of $R=3.2$.

The $\Lambda+\Sigma^0$ and $\Sigma^0$ directed flows are useful to determine
the repulsion of the $\Sigma$ single-particle potential,
which has larger theoretical uncertainty in their quasi-empirical values~\cite{Gal:2016boi}
compared to $\Lambda$ potential:
$U_\Lambda\,(\rho_0,k=0)\approx -30~\MeV$~\cite{Gal:2016boi,Jinno:2023xjr}
and $U_\Sigma\,(\rho_0,k=0)=30\pm 20~\MeV$~\cite{Gal:2016boi}.

\section{Summary}
\label{sec:summary}

We investigated the sensitivity of the $\Lambda$ and $\Sigma$
directed flows to the hyperon single-particle potential in
dense matter in the mid-central $\mathrm{Au}+\mathrm{Au}$
collision at a RHIC-BES energy, $\sqrt{s_{NN}}=4.5~\GeV$\@.
The single-particle potentials are obtained from
two- and three-body forces based on chiral EFT\@.
The resulting $\Sigma$ single-particle potential is
more repulsive compared to the $\Lambda$ one.
Their values at the saturation density are consistent
with the quasi-empirical values~\cite{Gal:2016boi}.

The $\Lambda+\Sigma^0$ directed flow is suppressed
when incorporating the $\Sigma$ single-particle potential.
The suppression is of the same magnitude as the uncertainty
in the momentum dependence.
This effect is more significant for the $\Sigma^0$ directed flow itself.
The $\Sigma$ potential has large uncertainty even at the saturation density
due to the lack of the $\Sigma$ hypernuclei
except for the special case ${}^4_\Sigma \mathrm{He}$.
The hyperon directed flows can help to further constrain properties
of the $\Sigma$ single-particle potential.

In this work, the hyperon resonances are assumed to feel the same
single-particle potentials as the corresponding ground state hyperons.
Their single-particle potentials should be investigated because
a number of resonances are produced during the early stage in the collisions
and affect the ground-state hyperon observables by the feed-down effect.
A work by using a parity-doublet model is ongoing and will be given elsewhere.

AJ would like to thank
Dominik Gerstung for kindly sharing his code with us.
This work was supported in part by
JST SPRING (No. JPMJSP2110)
and by the Grants-in-Aid for Scientific
Research from JSPS (%
	Nos. JP21K03577, 
	JP25K07284, 
	JP23K13102, 
	and JP25KJ1584). 

\bibliography{ref.bib}

\end{document}